\documentclass[12pt]{article}
\usepackage{amsmath}
\usepackage{amsfonts}
\usepackage{amssymb}
\usepackage{url}
\usepackage{hyperref}
\newcommand{\be}{\begin{equation}}
\newcommand{\ee}{\end{equation}}
\newcommand{\bea}{\begin{eqnarray}}
\newcommand{\eea}{\end{eqnarray}}

\begin{document}
\begin{titlepage}
\begin{flushright}
hep-ph/0608228
\end{flushright}
\vspace{4\baselineskip}
\begin{center}
{\Large\bf Axion and PVLAS data
in a Little Higgs model}
\end{center}
\vspace{1cm}
\begin{center}
{\large Takeshi Fukuyama$^{a,}$
\footnote{E-mail:fukuyama@se.ritsumei.ac.jp}}
and
{\large Tatsuru Kikuchi$^{b,}$
\footnote{E-mail:tatsuru@post.kek.jp}}
\end{center}
\vspace{0.2cm}
\begin{center}
${}^{a}$ {\small \it Department of Physics, Ritsumeikan University,
Kusatsu, Shiga, 525-8577, Japan}\\[.2cm]
${}^{b}$ {\small \it Theory Division, KEK,
Oho 1-1, Tsukuba, Ibaraki, 305-0801, Japan}\\
\medskip
\vskip 10mm
\end{center}
\vskip 10mm
\begin{abstract}
Little Higgs models may provide a solution to the gauge hierarchy
problem in the mass of the Higgs boson. In this framework 
the Higgs boson is arisen as the pseudo-Nambu-Goldstone (PNG) boson.
We show that the lepton triplet introduced in a little Higgs model
explains a small mass parameter in the double see-saw mechanism
for neutrino masses, and it can also gives an explanation for
the axion like particle recently reported by PVLAS collaboration.
\end{abstract}
\end{titlepage}
Despite a lot of phenomenological successes in the standard model (SM),
it has a fundamental problem associated with the Higgs mass parameter.
The Higgs mass squared parameter receives radiative corrections of order
the cutoff scale squared, implying the existence of some new physics
at a scale not much larger than the scale of electroweak symmetry breaking.
On the other hand, since experiments have not found any convincing sign of
such physics so far, it has been stimulated a lot of models of
the electroweak symmetry breaking.

Little Higgs models \cite{little} represent a new attempt to solve the gauge
hierarchy problem in the mass of the Higgs boson responsible for
the electroweak symmetry breaking.
This approach treats the Higgs boson as part of an assortment of
the pseudo-Nambu-Goldstone (PNG) bosons, arising from a spontaneously
broken global symmetry at a cutoff scale $\Lambda$, 
typically on the order of several TeV.

On the other hand, one and the most likely solution to the strong CP problem, 
Peccei-Quinn (PQ) solution gives us another interesting information 
to the physics beyond the SM \cite{PQ} that also predicts the existence of
a PNG, the so called axion. 
Axion particle $\phi$ has an interaction with photon as
\bea
{\cal L}&=&\frac{1}{4} g_{\phi \gamma \gamma} \phi
F_{\mu \nu} \widetilde{F}^{\mu \nu}
\nonumber\\
&=&
g_{\phi \gamma \gamma} \phi
\vec{E} \cdot \vec{B} \;.
\eea
The Primakov-like process allows $\phi \leftrightarrow \gamma$
conversion process to occur in the external $B$-field background.
The conversion rate for $\phi \to \gamma$ after traveling a distance $L$,
in the coherent limit, is given by
\be
P(\phi \to \gamma)=\frac{1}{4} g_{\phi \gamma \gamma}^2
B^2 L^2 \;.
\ee
The converted axion is reconverted to the photons
when it go across an optical barrier.
The detection rate for such photons in a second magnetic field
is given by \cite{VanBibber:1987rq}
\be
{\cal R} = P^2 \left(\frac{{\cal P}}{\omega} \right) \epsilon
= 0.6 \times 10^{17}~[{\rm sec^{-1}}]
\left(\frac{g_{\phi \gamma \gamma}}{10^{-6}~{\rm GeV^{-1}}} \right)^4
\left(\frac{B}{10~{\rm T}} \right)^4
\left(\frac{L}{10~{\rm m}} \right)^4
\epsilon\;,
\ee
where ${\cal P}$ is the laser power, $\omega$ is the photon energy and
$\epsilon$ is the efficiency for the detector.
This setup would be used in finding out the axion signal
at laser experiments.
Using the synchrotron X-rays from a free-electron laser (FEL)
with the parameters $B = 1$ T and $L = 1$ m, it can be probed for
the axion like particle with the mass range: $m_\phi \lesssim 10^{-3}$ eV,
and the coupling to the photon:
$g_{\phi \gamma \gamma} > 2.0 \times 10^{-8}~{\rm GeV^{-1}}$
within a year \cite{Rabadan:2005dm}.

The direct search for the production process 
$e^+ e^- \to \gamma^* \to \phi \gamma$
is given by the LEP experiment. 
The total cross section for such process is
\be
\sigma = 
g_{\phi \gamma \gamma}^2 \frac{\alpha (s - m_\phi^2)^2}{24 s^2} \;
\ee
and the decay rate for $\phi \to \gamma \gamma$ is given by
\be
\Gamma(\phi \to \gamma \gamma) 
= \frac{1}{64 \pi} g_{\phi \gamma \gamma}^2 m_\phi^3 \;.
\ee
The LEP data shows the following constraint:
\be
M_\phi=g_{\phi \gamma \gamma}^{-1} \gtrsim 10^5~{\rm GeV}\;.
\ee
%
In the usual PQ solution to the strong CP problem, we assume
an invisible axion with a decay constant $f_{a}$ that is
severely constrained from astrophysics to be
\be
10^{9 \pm 1} \; {\rm [GeV]} \lesssim f_a 
\lesssim 10^{12 \pm 1} \; {\rm [GeV]}\;.
\ee
Note the similarity of this scale and the right-handed neutrino
mass scale in the standard see-saw mechanism has been studied in
\cite{FK}.

Recently, PVLAS collabolation \cite{Zavattini:2005tm} reported the following
range of parameters for the axion-like particle, that is completely
different from the above range of parameters for the axion:
\bea
&&1 \times 10^5 ~{\rm GeV} < M_\phi < 6 \times 10^6~ {\rm GeV}\;,
\nonumber\\
&& 0.7 ~{\rm meV} < m_\phi < 2~ {\rm meV}\;.
\eea
As we will show in (18), the PVLAS data roughly equals to the scale of
symmetry breaking of order $\sqrt{m_\phi M_\phi} \sim 10~{\rm keV}$
unlike the QCD axion case: $\sqrt{m_a f_a} \sim \Lambda_{\rm QCD}
\simeq \sqrt{m_\pi f_\pi} \simeq 100~{\rm MeV}$.
It may be suggestive that some kind of anomaly may occur at 10 keV, 
though we do not discuss about its mechanism.

There have already been made several attempts \cite{PVLAS-modelA} to
explain the PVLAS data, and some models \cite{PVLAS-modelB} may explain
the apparent contradiction between PVLAS and CAST \cite{Zioutas:2004hi},
and PVLAS and the astrophysical bounds.

In this letter, we propose a little Higgs model, which can fit with
the PVLAS axion-like event.
The key ingredient for explaining the scale of $U(1)_{\rm PQ}$,
which is found to be of order $10~{\rm keV}$, in this letter
is to relate the scale with a mass parameter for the singlet
in the double see-saw mechanism \cite{Valle}
or the type-III see-saw mechanism \cite{type-III}
that gives an explanation for the smallness of neutrino masses.

We consider the simplest little Higgs model \cite{simplest}
and extend the model to include the Peccei-Quinn symmetry. 
The model is constructed to have Yukawa couplings and gauge 
interactions without introducing quadratic divergences.
It can be achieved by using the idea of ``collective symmetry breaking''. 
In order to realize this idea, we introduce $SU(3) \times SU(3)$
global symmetry which are spontaneously broken to $SU(2) \times SU(2)$ by
vacuum expectation values for two scalar triplets $\Phi_1$ and $\Phi_2$.
\be
\left<\Phi_1 \right> 
=
\frac{1}{\sqrt{2}}
\left(\begin{array}{c}
0 \\
0 \\
f
\end{array}\right)
\;,
~~
\left<\Phi_2 \right> 
=
\frac{1}{\sqrt{2}}
\left(\begin{array}{c}
0 \\
0 \\
f
\end{array}\right)\;.
\ee
For the collective symmetry breaking, the diagonal subgroup
$SU(3)_V = SU(3) \cap SU(3) \subset SU(3) \times SU(3)$ is gauged
and the subgroup $SU(2) \subset SU(3)_V$ will become the weak gauge
symmetry.
Using the residual $SU(3)_A \perp SU(3)_V$ symmetry, 
$\Phi_1$ and $\Phi_2$ are parametrized as
\bea
\Phi_1
&=&
\frac{1}{\sqrt{2}}
\exp \left[\frac{i}{f}
\left\{
\eta +
\left(\begin{array}{cc}
{\bf 0}_{2 \times 2} & h \\
h^\dag & 0
\end{array}\right)
\right\}
\right]
\left(
\begin{array}{c}
0 \\
0 \\
f 
\end{array}
\right) \;,
\nonumber\\
\Phi_2
&=&
\frac{1}{\sqrt{2}}
\exp \left[\frac{-i}{f}
\left\{
\eta +
\left(\begin{array}{cc}
{\bf 0}_{2 \times 2} & h \\
h^\dag & 0
\end{array}\right)
\right\}
\right]
\left(
\begin{array}{c}
0 \\
0 \\
f 
\end{array}
\right) \;.
\eea
In the little Higgs models, the Higgs mass parameters are protected
by global symmetries which include the Standard Model electroweak
gauge symmetry.
The tree level Lagrangian for the Higgs fields are only given
by the gauge interaction:
\be
{\cal L} = \left|(\partial_\mu  + ig A_\mu^a \lambda^a) \Phi_1 \right|^2+
\left|(\partial_\mu  + ig A_\mu^a \lambda^a) \Phi_2 \right|^2 \;,
\ee
where $\lambda^a~(a=1, \cdots, 8)$ is the Gell-Mann matrix.

After developing the VEVs for $\Phi_{1,2}$,
there appear extra gauge bosons $W^\prime$ and $Z^\prime$
whose masses are given by
\be
M_{W^\prime} = \frac{g}{\sqrt{2}} f \;,
~M_{Z^\prime} = \frac{\sqrt{2}g}{\sqrt{3 - \tan^2 \theta_W}} f\:.
\ee
The lower bound on $f$ is estimated to be
$f \gtrsim 4.5~{\rm TeV}$ \cite{Kilian:2003xt}.

The leading mass term for the Higgs boson is firstly generated
at the one-loop level
\be
M_{\rm Higgs}^2 \sim \frac{g^4}{16 \pi^2}
\log\left(\frac{\Lambda^2}{f^2} \right) f^2  
\sim \frac{g^4}{8 \pi^2}
\log\left(4 \pi \right) f^2  \;,
\ee
where we have used a relation $\Lambda \sim 4 \pi f$.
This $M_{\rm Higgs}$ is of the right order of magnitude to generate
the correct electroweak scale for $f \sim 10~{\rm TeV}$
and an order one quartic coupling $g$ is generated.

The Yukawa interactions for the matter multiplets should also be
extended in the little Higgs models. 
Here we consider the lepton sector \cite{delAguila:2005yi}.
The lepton doublet $L$ is enlarged to a triplet
${\bf 3} \sim \Psi=
\left(\begin{array}{c}
L \\
S
\end{array}\right)
$
by introducing a singlet $S$, and we also introduce a singlet
${\bf 1 } \sim N$ under any $SU(3)$ symmetries, which is necessary to have
a Yukawa coupling for $L$ and a triplet Higgs scalar $\Phi_{1}$:
\bea
-{\cal L}_Y
&=& y_\nu \overline{N} \Psi \Phi_1^\dag + h.c.
\nonumber\\
&=& -i y_\nu \overline{N} L h^\dag
+ y_\nu \overline{N} S \left(f - \frac{h^\dag h}{2f}\right) + h.c.
\eea
Under this Lagrangian, we can assign the PQ charges as listed in
the Table below.
\begin{center}
\begin{tabular}{|c|c|}
\hline \hline
fields & PQ charges \\
\hline
$\Phi_1$ & $+1$ \\
$\Phi_2$ & $-1$ \\
$\Psi $ & $-1$ \\
$N$ & $-2$ \\
\hline \hline
\end{tabular}
\end{center}
%
Since $S$ is a complete singlet under the SM, we can write down the Majorana
mass term for it:
\be
-{\cal L}_{\rm mass}
= \mu_s S^T C^{-1} S \;.
\ee
This term explicitly breaks PQ symmetry.
Then the total mass terms in the model can be written in a matrix form
in the base with $\{\nu,~N,~S \}$ as follows: 
\be
{\cal M} =
\left(
\begin{array}{ccc}
0    & m    &  0     \\
m    & 0    &  M \\
0    & M    &  \mu_s
\end{array}
\right)\;.
\label{double}
\ee
Here the mass parameter $m$ and $M$ are given by
\be
m =  -i y_\nu v, ~
M = y_{\nu} \left(f- \frac{v^2}{2f} \right) \; 
\ee
with $v = 174~{\rm GeV}$ represents the weak scale.
Hence, we obtain the light Majorana neutrino mass as
\be
m_\nu = \mu_s \left(\frac{m}{M} \right)^2
\sim
\mu_s \left(\frac{v^2}{f^2} \right)
\;.
\ee
For $m_\nu \sim 1~{\rm eV}$ and $f \sim 10~{\rm TeV}$, 
we have $\mu_s \sim 10~{\rm keV}$.

In summary, we have shown in this letter that the double see-saw mechanism has
its natural setup in the simplest little Higgs model. 
The little Higgs model indicates a Peccei-Quinn symmetry breaking around
$10~{\rm keV}$. It is very suggestive that this scale is the same order as
that of the scale for the axion-like particle recently reported by
the PVLAS collaboration.

\section*{Acknowledgments}
This work is supported in part by the Grant-in-Aid for Scientific Research 
from the Ministry of Education, Science and Culture of Japan (\#16540269).
We thank to the Theory Division at KEK for hospitality.
We are grateful to N. Gaur for his useful comments and discussions
and grateful to J. Reuter for his useful comments
on pseudo-axions in little Higgs models.



\begin{thebibliography}{99}
\bibitem{little}
N.~Arkani-Hamed, A.~G.~Cohen and H.~Georgi,
  Phys.\ Lett.\ B {\bf 513}, 232 (2001)
  [arXiv:hep-ph/0105239];
%
N.~Arkani-Hamed, A.~G.~Cohen, E.~Katz and A.~E.~Nelson,
  JHEP {\bf 0207}, 034 (2002)
  [arXiv:hep-ph/0206021];
%
N.~Arkani-Hamed, A.~G.~Cohen, E.~Katz, A.~E.~Nelson, T.~Gregoire and J.~G.~Wacker,
  JHEP {\bf 0208}, 021 (2002)
  [arXiv:hep-ph/0206020];
%
T.~Gregoire and J.~G.~Wacker,
  JHEP {\bf 0208}, 019 (2002)
  [arXiv:hep-ph/0206023];
%
I.~Low, W.~Skiba and D.~Smith,
  Phys.\ Rev.\ D {\bf 66}, 072001 (2002)
  [arXiv:hep-ph/0207243];
%
D.~E.~Kaplan and M.~Schmaltz,
  JHEP {\bf 0310}, 039 (2003)
  [arXiv:hep-ph/0302049];
%
S.~Chang and J.~G.~Wacker,
  Phys.\ Rev.\ D {\bf 69}, 035002 (2004)
  [arXiv:hep-ph/0303001];
%
W.~Skiba and J.~Terning,
  Phys.\ Rev.\ D {\bf 68}, 075001 (2003)
  [arXiv:hep-ph/0305302];
%
S.~Chang,
  JHEP {\bf 0312}, 057 (2003)
  [arXiv:hep-ph/0306034];
%
H.~C.~Cheng and I.~Low,
  JHEP {\bf 0309}, 051 (2003)
  [arXiv:hep-ph/0308199];
%
  JHEP {\bf 0408}, 061 (2004)
  [arXiv:hep-ph/0405243];
%
M.~Schmaltz,
  JHEP {\bf 0408}, 056 (2004)
  [arXiv:hep-ph/0407143].

\bibitem{PQ}
R.~D.~Peccei, H.~R.~Quinn, 
Phys.\ Rev.\ Lett. {\bf 38}, 1440 (1977); 
Phys.\ Rev. D {\bf 16}, 1791 (1977).

\bibitem{VanBibber:1987rq}
K.~VanBibber, N.~R.~Dagdeviren, S.~E.~Koonin, A.~K.~Kerman, H.~N.~Nelson,
  Phys.\ Rev.\ Lett.\  {\bf 59}, 759 (1987);
  G.~Ruoso {\it et al.},
  Z.\ Phys.\ C {\bf 56}, 505 (1992);
  R.~Cameron {\it et al.},
  Phys.\ Rev.\ D {\bf 47}, 3707 (1993).

\bibitem{Rabadan:2005dm}
  R.~Rabadan, A.~Ringwald and K.~Sigurdson,
  Phys.\ Rev.\ Lett.\  {\bf 96}, 110407 (2006)
  [arXiv:hep-ph/0511103].

\bibitem{FK}
  T.~Fukuyama and T.~Kikuchi,
  JHEP {\bf 0505}, 017 (2005)
  [arXiv:hep-ph/0412373].

\bibitem{Zavattini:2005tm}
  E.~Zavattini {\it et al.}  [PVLAS Collaboration],
  Phys.\ Rev.\ Lett.\  {\bf 96}, 110406 (2006)
  [arXiv:hep-ex/0507107].

\bibitem{PVLAS-modelA}
E.~Masso and R.~Toldra,
  Phys.\ Rev.\ D {\bf 52}, 1755 (1995)
  [arXiv:hep-ph/9503293];
%
M.~Kleban and R.~Rabadan,
  arXiv:hep-ph/0510183;
%
A.~Ringwald,
  J.\ Phys.\ Conf.\ Ser.\  {\bf 39}, 197 (2006)
  [arXiv:hep-ph/0511184];
%
C.~Biggio, E.~Masso and J.~Redondo,
  arXiv:hep-ph/0604062;
%
J.~P.~Conlon,
  arXiv:hep-ph/0607138.

\bibitem{PVLAS-modelB}
E.~Masso and J.~Redondo,
  JCAP {\bf 0509}, 015 (2005)
  [arXiv:hep-ph/0504202];
%
  E.~Gabrielli, K.~Huitu and S.~Roy,
  arXiv:hep-ph/0604143;
%
I.~Antoniadis, A.~Boyarsky and O.~Ruchayskiy,
  arXiv:hep-ph/0606306;
%
J.~Jaeckel, E.~Masso, J.~Redondo, A.~Ringwald and F.~Takahashi,
  arXiv:hep-ph/0605313;
E.~Masso and J.~Redondo,
  arXiv:hep-ph/0606163;
  arXiv:hep-ph/0606164;
%
E.~Masso,
  arXiv:hep-ph/0607215.

\bibitem{Zioutas:2004hi}
  K.~Zioutas {\it et al.}  [CAST Collaboration],
  Phys.\ Rev.\ Lett.\  {\bf 94}, 121301 (2005)
  [arXiv:hep-ex/0411033].

\bibitem{Valle}
R.~N.~Mohapatra,
Phys.\ Rev.\ Lett.\  {\bf 56} (1986) 561;
R.~N.~Mohapatra and J.~W.~F.~Valle,
Phys.\ Rev.\ D {\bf 34}, 1642 (1986);
T.~Fukuyama, A.~Ilakovac, T.~Kikuchi and K.~Matsuda,
  JHEP {\bf 0506}, 016 (2005)
  [arXiv:hep-ph/0503114].

\bibitem{type-III}
S.~M.~Barr,
Phys.\ Rev.\ Lett.\  {\bf 92}, 101601 (2004)
[arXiv:hep-ph/0309152].

\bibitem{simplest}
M.~Schmaltz,
  Nucl.\ Phys.\ Proc.\ Suppl.\  {\bf 117}, 40 (2003)
  [arXiv:hep-ph/0210415];
D.~E.~Kaplan and M.~Schmaltz,
  JHEP {\bf 0310}, 039 (2003)
  [arXiv:hep-ph/0302049].

\bibitem{Kilian:2003xt}
  W.~Kilian and J.~Reuter,
  Phys.\ Rev.\ D {\bf 70}, 015004 (2004)
  [arXiv:hep-ph/0311095];
J.~A.~Casas, J.~R.~Espinosa and I.~Hidalgo,
  JHEP {\bf 0503}, 038 (2005)
  [arXiv:hep-ph/0502066];
G.~Marandella, C.~Schappacher and A.~Strumia,
  Phys.\ Rev.\ D {\bf 72}, 035014 (2005)
  [arXiv:hep-ph/0502096];
  Z.~Han and W.~Skiba,
  Phys.\ Rev.\ D {\bf 72}, 035005 (2005)
  [arXiv:hep-ph/0506206].

\bibitem{delAguila:2005yi}
F.~del Aguila, M.~Masip and J.~L.~Padilla,
  Phys.\ Lett.\ B {\bf 627}, 131 (2005)
  [arXiv:hep-ph/0506063];
A.~Abada, G.~Bhattacharyya and M.~Losada,
  Phys.\ Rev.\ D {\bf 73}, 033006 (2006)
  [arXiv:hep-ph/0511275].

\end{thebibliography}
\end{document}